\documentclass[11pt]{article}
\usepackage{amstext}
\usepackage{amsfonts}
\usepackage{amsmath}

\newtheorem{theorem}{Theorem}[section]

\newtheorem{proposition}[theorem]{Proposition}

\newtheorem{corollary}[theorem]{Corollary}
\newtheorem{definition}{Definition}[section]

\newtheorem{remark}[theorem]{Remark}

\def\square{\rule{2mm}{2mm}}
\newenvironment{proof}{{\noindent\bf Proof:  }}{\qquad\square}

\def\squarebox#1{\hbox to #1{\hfill\vbox to #1{\vfill}}}

\newcommand{\tensor}{\otimes}
\newcommand{\Tensor}{\bigotimes}
\newcommand{\xor}{\oplus}

\newcommand\meet\wedge

\newcommand{\adjoint}{\dagger}
\newcommand{\integral}{\int}

\newcommand{\complex}{{\mathbb C}}
\newcommand{\reals}{{\mathbb R}}
\newcommand{\GF}{{\mathrm{GF}}}

\newcommand{\complexi}{\mathrm{i}}
\newcommand{\e}{\mathrm{e}}

\newcommand{\trace}{{\rm Tr}}
\newcommand{\expct}{{\mathrm E}}
\newcommand{\size}[1]{\left|#1\right|}

\newcommand{\ket}[1]{|#1\rangle}
\newcommand{\bra}[1]{\langle #1|}

\newcommand{\ketbra}[2]{\ket{#1}\!\bra{#2}}
\newcommand{\density}[1]{\ketbra{#1}{#1}}
\newcommand{\norm}[1]{\left\|\,#1\,\right\|}
\newcommand{\trnorm}[1]{\norm{#1}_{\mathrm {tr}}}
\newcommand{\set}[1]{{\left\{#1\right\}}}

\newcommand{\dotprod}[1]{{\left\langle #1 \right\rangle}}

\newcommand{\Order}{{\mathrm{O}}}

\newcommand{\tOrder}{{\mathrm{\tilde{O}}}}

\newcommand{\identity}{{\mathbb I}}
\newcommand{\linear}{{\mathrm{L}}}
\newcommand{\unitary}{{\mathrm{U}}}
\newcommand{\paulix}{{\mathrm X}}
\newcommand{\pauliy}{{\mathrm Y}}
\newcommand{\pauliz}{{\mathrm Z}}
\newcommand{\paulii}{{\mathrm I}}
\newcommand{\pauli}{{\mathrm P}}

\newcommand{\ignore}[1]{}
\newcommand{\suppress}[1]{}

\newcommand{\etal}{{\it et al\/}}

\newcommand{\aitch}{{\mathcal{H}}}
\newcommand{\emm}{{\mathcal{M}}}
\newcommand{\bias}{{\mathrm{bias}}}

\begin{document}

\title{\bf Approximate Randomization of Quantum States With Fewer Bits
of Key}

\author{
Paul A.\ Dickinson~\thanks{
Department of Combinatorics and Optimization, and
Institute for Quantum Computing,
University of Waterloo.
Address:
200 University Ave.\ W., Waterloo, Ontario N2L 3G1, Canada.
Email:  {\tt padickinson@iqc.ca}.
Research supported in part by an NSERC Canada Graduate Scholarship,
and grants from NSERC Canada, CIAR, MITACS, CFI, and OIT (Canada).} \\
University of Waterloo
\and
Ashwin Nayak~\thanks{
Department of Combinatorics and Optimization,
and Institute for Quantum Computing, University of Waterloo, 
and Perimeter Institute for Theoretical Physics.
Address: 200 University Ave.\ W., Waterloo, Ontario N2L 3G1, Canada.
Email: {\tt anayak@math.uwaterloo.ca}.
Research supported in part by NSERC Canada, CIAR, MITACS, CFI, and OIT
(Canada). Research at Perimeter Institute is supported in part by the
Government of Canada through NSERC and by the Province of Ontario
through MEDT.
} \\
University of Waterloo, and \\
Perimeter Institute
}

\date{June 1, 2006}

\maketitle

\begin{abstract}
Randomization of quantum states is the quantum analogue of
the classical one-time pad.  We present an improved, efficient 
construction of an approximately randomizing map that
uses~$\Order(d/\epsilon^2)$
Pauli operators to map any~$d$-dimensional state to a state that is
within trace distance~$\epsilon$ of the completely mixed state.
Our bound is a~$\log d$ factor smaller than that of Hayden, Leung,
Shor, and Winter~\cite{HaydenLSW04}, and Ambainis and 
Smith~\cite{AmbainisS04}.

Then, we show that a random sequence of essentially the same 
number of unitary operators, chosen from an appropriate set, with high 
probability form an approximately randomizing map for~$d$-dimensional states.
Finally, we discuss the optimality of these schemes via connections to 
different notions of pseudorandomness, and give a new lower bound for
small~$\epsilon$.
\end{abstract}

\section{Introduction}
\label{sec-introduction}

\subsection{Encryption of quantum states}

Randomization of quantum states is a procedure analogous to
encryption of classical messages such as in the ``one-time pad''. 
Imagine that two parties wish to exchange sensitive data in the
form of quantum states over an insecure quantum communication channel.
They would like to encrypt the quantum data so that any eavesdropper
with access to the channel will not gain any information 
about the
data.  The idea is to use a secret key, such as a uniformly random bit
string, to transform a
quantum state so that without access to the key, an adversary is
unable to distinguish two different encrypted states,
when averaged over the random choice of key.
Equivalently, every state is mapped to the same mixed
state by the encryption procedure. The encrypted state may then be sent
over the insecure channel, and the receiver, who also knows the key may
decrypt to recover the message. 

It has been known for several years that applying an independently chosen 
random Pauli operator to each qubit of an~$n$-qubit state maps it to the
completely mixed state~$\identity/2^n$. This gives rise to a scheme for
perfect encryption of~$n$-qubit quantum states with~$2n$ secret uniformly
random classical bits~\cite{BoykinR03,AmbainisMTW00}. This was also
shown to be optimal in terms of the number of bits of key 
required~\cite{BoykinR03,AmbainisMTW00,Jain05,NayakSen06}.

The requirement of perfect encryption may be relaxed a little without 
compromising security, so that the encrypted states are all {\em close\/}
to being completely mixed, rather than being exactly so. By using a
probabilistic argument, Hayden, Leung,
Shor, and Winter~\cite{HaydenLSW04} showed that the number of bits of
key required then drops by a factor of approximately~$2$: to
approximately randomize~$n$-qubit states to within~$\epsilon$
of~$\identity/2^n$ (in trace norm), we need at most~$n + \log n + 2\log 
\tfrac{1}{\epsilon} + \Order(1)$ bits of key. Subsequently, 
Ambainis and Smith~\cite{AmbainisS04} gave an efficient (quadratic time)
scheme for 
approximate state randomization with respect to the trace norm using
\[
n + \min \set{2 \log n + 2\log \frac{1}{\epsilon}, \log n + 3 \log
\frac{1}{\epsilon} } + \Order(1)
\]
bits of key.
Their construction is based on {\em small-bias spaces\/} (see, e.g.,
Ref.~\cite{NaorN93}).
They also showed how to reduce the key length to~$n + 2\log
\tfrac{1}{\epsilon}$ at the cost of
increasing the length of the ciphertext by~$2n$ bits.

The amount of key required for approximate encryption with respect 
to the Hilbert-Schmidt norm and the operator norm has been studied by 
Kerenidis and Nagaj~\cite{KerenidisN05}. They show that key length is
quite sensitive to the norm chosen to specify the security requirement.

In this article, we revisit
approximate randomization with respect to the trace norm, which 
reflects most closely our ability to physically distinguish quantum
states.
We first observe that an explicit scheme of Ambainis and Smith
may be improved by using an optimal construction of 
small-bias spaces due to Alon,
Goldreich, H{\aa}stad, and Peralta~\cite{AlonGHP92}. This 
reduces the key size to~$n + 2 \log \tfrac{1}{\epsilon} + 4$, and
avoids the need for ciphertext that 
is longer than the original message. This construction avoids another 
rather subtle issue. The length-preserving schemes suggested in
Ref.~\cite{AmbainisS04} require that the two communicating
parties agree on a prime number of length~$\Theta(n)$. Since there is no
known polynomial-time
deterministic procedure to generate a prime number of a specified
length, additional communication is required to establish this shared
prime number. (The prime may be generated locally by one party by a 
randomized procedure.) The encryption and decryption procedures we
suggest require a common irreducible polynomial 
over~$\GF(2)$ of degree~$\Theta(n)$, 
which may be computed independently by the two parties using an
efficient deterministic algorithm due to Shoup~\cite{Shoup89}.

Next, we investigate collections of unitary operators that give rise to
approximately randomizing maps.  We show by a probabilistic argument that
any sequence of
\[
\Order\left( \frac{d}{\epsilon^2} \log\frac{1}{\epsilon} \right)
\]
unitary operators chosen independently from a perfectly randomizing set
with high probability defines an approximately randomizing map
for~$d$-dimensional quantum states. 

A simple rank argument shows that at least~$d(1 - \tfrac{\epsilon}{2})$
unitary operators are always needed, for approximate encryption in~$d$
dimensions. No better lower bound is known.
Methods for showing lower bounds for perfect encryption all fail, since
they crucially 
rely on the property of completely randomizing maps to destroy all
quantum correlation between the encrypted state and any state previously 
entangled to it. We take a different approach, and
derive necessary conditions on distributions over
Pauli operators that correspond to approximately randomizing maps. These
conditions are similar to notions such as
``almost $n$-wise independence'' in the theory of pseudorandom
distributions (see, e.g., Ref.~\cite{NaorN93}). As a corollary, we get a
tighter lower bound on randomizing sets of Pauli operators in the regime
of exponentially small~$\epsilon$.

We describe our results more formally in the following two subsections, and
then prove them in the remaining sections.

\subsection{Preliminaries}
\label{sec-prelim}

We refer the reader to the text~\cite{NielsenC00} or the lecture
notes~\cite{Preskill98} for definitions of basic concepts in quantum
information.

Let~$\linear(\aitch)$ denote the space of linear operators on the
Hilbert space~$\aitch$.  This includes the cone of positive
semi-definite operators (density operators)
on~$\aitch$. Let~$\unitary(\aitch)$ denote the set of unitary linear
operators on the Hilbert space~$\aitch$.

\begin{definition}
Let~$\epsilon \geq 0$.
A completely positive, trace-preserving (CPTP) linear operator~$R :
\linear(\complex^d) \rightarrow \linear(\complex^d)$ is said to
be~$\epsilon$-randomizing with respect to the norm~$\norm{\cdot}$ if,
for all density operators (mixed states)~$\rho \in
\linear(\complex^d)$,
\begin{eqnarray*}
\norm{ R(\rho) - \frac{\identity}{d} } & \le & \epsilon.
\end{eqnarray*}
We say that~$R$ is {\it completely\/} randomizing if~$\epsilon = 0$.
\end{definition}
\begin{remark}
\label{rem-pure}
Due to convexity, a map~$R$ that randomizes all pure states (rank~$1$
density operators) also randomizes all {\em mixed\/} states to the
same extent. 
\end{remark}

We will mainly discuss randomization with respect to the trace
norm. For any linear operator~$M \in \linear(\complex^d)$, the trace
norm is defined by~$\trnorm{M} = \trace \sqrt{M^\adjoint
M}$. Equivalently, it is the sum of the singular values of~$M$, and
therefore also referred to as the~``$1$-norm''.  The trace norm is
arguably a more appropriate measure of distinguishability in the
context of eavesdropping, since it is directly related to information
that measurements reveal about quantum states.  We will also use the
Frobenius (or Hilbert-Schmidt) norm in our proofs. This norm is
defined as~$\norm{M}_F = \sqrt{\trace(M^\adjoint M)}$. Since this is
the~$\ell_2$ norm of the vector of singular values of~$M$, this is
also referred to as the~``$2$-norm'' by some authors.

Randomizing maps are easy to construct. For example, the map~$R : \rho
\mapsto \trace(\rho) \frac{\identity}{d}$ is completely
randomizing. However, these maps are most useful when they can be
inverted by a quantum operation to recover the original state, as is
required in the case of encryption.

The protocols for encryption we study involve two parties, labeled
Alice and Bob, who share a secret, uniformly random bit-string, called
the private key~$k$. Alice wishes to send a~$d$-dimensional quantum
state~$\rho$ to Bob. She would like to apply an invertible quantum
operation~\cite{NayakSen06} to the state, and send it to Bob so that
when averaged over~$k$, the map is randomizing. This would ensure that
no eavesdropper be able to distinguish two different messages with
non-trivial probability. Such protocols have also been called
``private quantum channels'' by some authors (see, e.g.,
Ref.~\cite{AmbainisMTW00}).  We have implicitly assumed that the
quantum channel is noiseless unless an eavesdropper tampers with
it. Therefore Bob, who also has the key~$k$, can apply the inverse
operation to decrypt the message~$\rho$ perfectly.

A natural way to create such an invertible randomizing operator is to
select a sequence of unitary operators~$U_1, \ldots, U_m$ and define
\begin{eqnarray}
\label{eqn-randmap}
R(\rho) & = & \frac{1}{m} \sum_{i = 1}^m U_i \rho U_i^\adjoint.
\end{eqnarray}
Here, the index~$i$ corresponds to the shared secret key held by the
communicating parties, and is unknown to any eavesdropper. With a
suitable choice of unitary operators the map~$R$ would
be~$\epsilon$-randomizing. In fact, any orthogonal set of unitary
operations on~$\complex^d$, such as the set of~$d^2$ Pauli operators,
form a completely randomizing map~\cite{BoykinR03}.

The most general one-way encryption scheme may in addition involve an
ancilla that depends upon the key~\cite{NayakSen06}:
\begin{eqnarray*}
R(\rho) & = & \frac{1}{m} \sum_{i = 1}^m U_i (\rho \tensor \sigma_i)
              U_i^\adjoint.
\end{eqnarray*}
This is slightly more general than the form claimed in
Ref.~\cite{AmbainisMTW00}. However, the results in the latter article
extend to the more general maps above (see also
Ref.~\cite{Jain05}). This general form of encryption uses more qubits
in the ciphertext than originally present in the message, which is
undesirable from an efficiency point of view. We will only study
randomizing maps as in equation~(\ref{eqn-randmap}), which correspond
to encryption without ancilla.

The randomizing maps we construct will involve the Pauli operators.
We will denote the Pauli operators on a single qubit by~$\paulii, \paulix,
\pauliy, \pauliz$:
\[
\paulii  =  \left(\begin{array}{cc} 1 & 0 \\ 0 & 1 \end{array} \right),\quad
\paulix  =  \left(\begin{array}{cc} 0 & 1 \\ 1 & 0 \end{array} \right), \quad
\pauliz  =  \left(\begin{array}{cr} 1 & 0 \\ 0 & -1 \end{array} \right), \quad
\pauliy  = \complexi \paulix\pauliz 
= \left(\begin{array}{cr} 0 & -\complexi \\ \complexi & 0 \end{array} \right).
\]
These operators are unitary, Hermitian, and they square to the
identity. The non-identity Pauli matrices anti-commute with other
non-identity Pauli matrices. For example,~$\paulix \pauliy = - \pauliy
\paulix$.  Where the overall phase of~`$\complexi$' is irrelevant, we
will substitute~$\pauliy$ with the matrix~$\paulix\pauliz$.

For two~$n$-bit strings~$a,b$, let~$\size{a \meet b} = \sum_{j = 1}^n
a_j b_j$.  We will often represent a tensor product of~$n$ single
qubit Pauli operators by a string of~$2n$ bits~$(a,b) \in
\set{0,1}^{2n}$ using the correspondence
\begin{eqnarray}
\label{eqn-bijection}
(a,b) & \leftrightarrow &  \complexi^{\size{a \meet b}} \,
    \paulix^a \pauliz^b, \quad \text{where} \\
 \nonumber
\paulix^a & = & \paulix^{a_1} \tensor \paulix^{a_2} \tensor 
                \cdots \tensor \paulix^{a_n},
\end{eqnarray}
and~$\pauliz^b$ is defined similarly. Let~$\pauli_n$ denote the
set~$\set{ \complexi^{\size{a \meet b}} \paulix^a \pauliz^b : (a,b)
\in \set{0,1}^{2n} }$ of all tensor products of~$n$ single qubit Pauli
operators.

For two~$n$-bit strings~$a,b \in \set{0,1}^n$, considered as elements
of~$\GF(2)^n$, the standard scalar product is defined
as~$\dotprod{u,x} = \sum_i u_i x_i \pmod{2}$. The symplectic inner
product of a pair of~$2n$-bit strings~$(a,b)$ and~$(c,d)$, considered
as elements of~$\GF(2)^{2n}$, is given by~$\dotprod{a,d} +
\dotprod{b,c} \pmod{2}$. The symplectic inner product tells us when
two Pauli operators commute: $\paulix^a \pauliz^b$ commutes
with~$\paulix^c \pauliz^d$ if and only if the symplectic inner product
of~$(a,b)$ and~$(c,d)$ is~$0$.

A distribution~$p$ over~$\set{0,1}^{2n}$ defines a CPTP map on~$n$
qubits via the above bijection:
\begin{eqnarray}
\label{eqn-paulimap}
R_p(\rho) & = & \sum_{(a,b)\in \set{0,1}^{2n}} p(a,b) \; \paulix^a
              \pauliz^b \rho \pauliz^b \paulix^a.
\end{eqnarray}
We will study randomizing maps of this form more closely. A special
case is when~$p$ is uniformly distributed over a set~$S \subset
\set{0,1}^{2n}$. In this case, we will denote the associated CPTP map
by~$R_S$.

The single qubit Pauli operators~$\pauli_1$ form an orthogonal basis
for~$\linear(\complex^2)$ under the inner product~$(A,B) =
\trace(B^\adjoint A)$.  The set~~$\pauli_n$ of~$2^{2n}$ tensor
products of~$n$ such Pauli operators similarly form an orthogonal
basis for~$\linear(\complex^{2^n})$.  There are also bases of~$d^2$
orthogonal unitary operators on~$\linear(\complex^d)$ for general
dimension~$d$ (that is not a power of~$2$).

\suppress{ Let~$\pauli_n$ denote the group
\[
\set{ \alpha \Tensor_{i = 1}^n M_i : \alpha \in \set{\pm 1, \pm \complexi},
M_i \in \set{\identity, \paulix, \pauliy, \pauliz}}, 
\]
of Pauli operators on~$n$ qubits.
}

We will also make use of the concept of a {\em stabilizer
state\/}~\cite[Section~10.5.1, page~454]{NielsenC00}. A {\em
stabilizer group\/}~$G$ is an abelian group generated by a subset~$T
\subset \pauli_n$ of the Pauli operators on~$n$ qubits.  Each
stabilizer group~$G$ defines a linear subspace~$C_G$
of~$\complex^{2^{n}}$ which is the common~$+1$-eigenspace of all the
Pauli operators in~$G$. If~$G$ is generated by~$k$ independent Pauli
operators, and does not contain~$-\identity$, then the linear
subspace~$C_G$ has dimension~$2^{n-k}$. By a stabilizer state, we will
mean a pure state which spans the one dimensional subspace~$C_G$
stabilized by a group~$G$ of order~$2^n$.

Every stabilizer group generated by~$k$ independent Pauli matrices may
be specified by listing its generators row-wise in a~$k \times 2n$
boolean matrix~$M$ via the bijection in
equation~(\ref{eqn-bijection}).  Since the generators all commute,
different rows of the matrix have symplectic inner product~$0$ with
each other. For a $2n$-bit vector~$w = (u,v)$, let~$M \cdot w$ denote
the $k$-bit vector obtained by taking the symplectic inner product of
the~$k$ rows of~$M$ with~$w$.

\subsection{Statement of Results}
\label{sec-results}

The problem we address in this paper is the construction of
approximately randomizing maps which preserve the number of qubits in
the message.

First, in Section~\ref{sec-construction}, we describe an an explicit
construction for a sequence of unitaries that approximately
randomize. This construction combines the work of
Refs.~\cite{AmbainisS04,AlonGHP92,Shoup89} to give an improvement over
the explicit construction by Ambainis and Smith~\cite{AmbainisS04}.

\begin{theorem}
\label{thm-constructive}
For any~$\epsilon \in (0,2]$, and dimension~$d = 2^n$, there is a
sequence of~$m = \frac{16d}{\epsilon^2}$ unitary operations~$\set{U_i
: 1 \le i \le m}$, each a tensor product of Pauli operators, such that
the map
\begin{eqnarray*}
R(\rho) & = & \frac{1}{m} \sum_{i = 1}^m U_i \rho U_i^\adjoint
\end{eqnarray*}
is~$\epsilon$-randomizing with respect to the trace norm.  The
sequence of Pauli operators defining~$U_i$ may be determined from the
index~$i$ in time~$\tOrder((\log m)^4) = \tOrder(n^4)$.
\end{theorem}
\begin{remark}
The notation~$\tOrder(T)$ above suppresses factors poly-logarithmic
in~$T$. Since there is an linear-time completely randomizing map
consisting of a sequence of~$d^2$ unitary operators, the above theorem
is only useful when~$\epsilon > 4/\sqrt{d} = 4/2^{n/2}$. We were
therefore able to assume that~$\log m = \log d + 2 \log
\tfrac{1}{\epsilon} + \Order(1) = \Order(n)$.
\end{remark}

Next, we study which sequences of unitary operations are suitable for
approximate encryption.  In Section~\ref{sec-random}, we prove that
almost all sequences of~$\Order(\tfrac{d}{\epsilon^2}
\ln\tfrac{1}{\epsilon})$ unitary operations form an~$\epsilon$
approximately randomizing map for~$d$ dimensional states.

\begin{theorem}
\label{thm-randomized}
For all~$\epsilon \in (0,2]$, a random sequence of~$m =
\frac{37d}{\epsilon^2} \ln \left( \frac{15}{\epsilon} \right)$ unitary
operations~$\set{U_i : 1 \le i \le m}$ in~$\unitary(\complex^d)$
defines a map
\begin{eqnarray*}
R(\rho) & = & \frac{1}{m} \sum_{i = 1}^m U_i \rho U_i^\adjoint
\end{eqnarray*}
and~$R$ is~$\epsilon$-randomizing with respect to the trace norm, with
probability at least $1-\e^{-d/2}$. Each unitary operation~$U_i$ may be
chosen independently from an arbitrary distribution
over~$\unitary(\complex^d)$ that gives rise to a completely randomizing
map.
\end{theorem}
In the above theorem, each unitary in the sequence may be chosen
independently according to an arbitrary completely randomizing
distribution of unitaries, not necessarily the same for each~$i$. For
instance, it may be chosen according to the Haar measure
on~$\unitary(\complex^d)$, or the uniform distribution over any
orthogonal unitary basis for~$\unitary(\complex^d)$.  For the case
that~$\rho$ is an $n$-qubit state, the unitary operators can be chosen
from among the Pauli operators, which are particularly simple
operators.

Theorem~\ref{thm-randomized} is in general incomparable to
Theorem~II.2 of Hayden \etal.~\cite{HaydenLSW04}. Our theorem reduces
by a factor of~$\log d$ the number of unitaries required for
approximate encryption in the trace norm. However, it does not imply
the stronger bound of~$\epsilon/d$ with respect to the spectral norm
(``$\infty$-norm'') on the distance from the completely mixed state,
even with a~$\log d$ factor more unitaries.

We conjecture that the construction of the approximately randomizing
map in Theorem~\ref{thm-constructive} is optimal in the use of secret
key bits, up to an additive constant. We are unable to establish this
rigorously at present, but take some steps towards this.

We derive conditions on distributions over~$\set{0,1}^{2n}$ that
define randomizing maps. These conditions are similar in flavour to
other notions of pseudo-randomness such as ``almost $k$-wise
independence''.  We believe these will help prove the optimality of
our constructions.

\begin{theorem}
\label{thm-constraints}
Let~$R_p$ be a CPTP map on~$n$ qubits induced by a distribution~$p$
over~$\set{0,1}^{2n}$, as in equation~(\ref{eqn-paulimap}). Let~$V$ be
the random variable corresponding to~$p$. If~$R_p$ is an
$\epsilon$-randomizing map with respect to the trace norm, then the
random variable~$M \cdot V$ is $\epsilon$-close to the uniform
distribution over~$\set{0,1}^n$ in~$\ell_1$ distance for every~$n
\times 2n$ matrix~$M$ over~$\GF(2)$ that defines a stabilizer state.
\end{theorem}

As a corollary, we prove that any distribution corresponding to
an~$\epsilon$ randomizing map (with respect to the trace norm) is
necessarily~$\epsilon$-biased (cf.\ Definition~\ref{def-bias} in
Section~\ref{sec-construction}). This implies a new lower bound on the
number of bits of key in the regime of extremely small~$\epsilon$,
when it is smaller than~$2^{-n/2}$.

\begin{corollary}
\label{thm-converse}
Let~$R_p$ be a CPTP map on~$n$ qubits induced by a distribution~$p$
over~$\set{0,1}^{2n}$, as in equation~(\ref{eqn-paulimap}).  If~$R_p$
is an $\epsilon$-randomizing map with respect to the trace norm, then
the distribution~$p$ is $\epsilon$-biased.  Therefore, if~$p$ 
has support~$S \subset \set{0,1}^{2n}$, then~$\size{S}$ is at least a
universal constant times
\[
\min \set{ 2^{2n}, \quad \frac{n}{\epsilon^2 \log\frac{1}{\epsilon}} }.
\]
\end{corollary}

\section{An explicit randomizing set}
\label{sec-construction}

In this section, we prove Theorem~\ref{thm-constructive}. We describe
an explicit sequence of unitary (Pauli) operators that are
approximately randomizing. The~$i$-th unitary in the sequence can be
determined from the index~$i$ in polynomial time. To obtain this
result, we use the connection made by Ambainis and
Smith~\cite{AmbainisS04} between randomizing maps and {\em small-bias
spaces\/}, together with a more efficient construction of such spaces
due to Alon, Goldreich, H{\aa}stad, and Peralta~\cite{AlonGHP92}.

Recall that the Pauli operators form an orthogonal basis for matrices,
so we may express any density matrix over~$n$ qubits as
\begin{eqnarray*}
\rho & = & \sum_{M \in \pauli_n}
           \frac{\trace(M^\adjoint\rho)}{\trace(M^\adjoint M)} \; M \\
     & = & \frac{1}{2^n} \sum_{M \in \pauli_n} \alpha_M \; M,
\end{eqnarray*}
where~$\alpha = (\alpha_M)$ is a vector in~$\complex^{2n}$
with~$\norm{\alpha}_2^2 \leq 2^n$.  The
component~$\alpha_\identity/2^n$ of any quantum state along the
identity operator is exactly~$1/d = 1/2^n$. If a CPTP map~$E$ is
completely randomizing, then
\begin{eqnarray*}
E(\rho) & = & \frac{1}{2^n} \sum_{M \in \pauli_n} \alpha_M \; E(M) \\
        & = & \frac{1}{2^n} \paulii.
\end{eqnarray*}
Thus, the map annihilates all the non-identity components of the
state. The idea behind the construction for approximate randomization
is to construct a map that shrinks the non-identity components of a
density matrix sufficiently, so that it becomes close to completely
mixed. Such a map may be constructed from {\em small-bias\/} sets.

\begin{definition}[Naor and Naor~\cite{NaorN93}]
\label{def-bias}
The {\em bias\/} of a subset~$S \subset \set{0,1}^k$ with respect to a
string~$u \in \set{0,1}^k$ is defined as
\begin{eqnarray*}
\bias(S,u)
    & = & \size{ \expct_{x \in S} \; (-1)^\dotprod{u,x} } \\
    & = & \size{ 1 - 2 \; \expct_{x \in S} \; \dotprod{u,x} },
\end{eqnarray*}
where the expectation is taken over strings~$x$ chosen uniformly at
random from~$S$, and $\dotprod{u,x} = \sum_i u_i x_i
\pmod{2}$ is the standard scalar product over~$\GF(2)$.

The subset~$S \subset \{0,1\}^k$ is said to be {\em $\delta$-biased\/}
if the bias with respect to every non-zero string is bounded
by~$\delta$: $\bias(S,u) \le \delta$ for all~$u \in
\set{0,1}^k - \set{0^k}$.

This definition extends to arbitrary distributions~$p$
over~$\set{0,1}^k$ in the natural way: $\bias(p,u) = \size{ \expct \;
(-1)^\dotprod{u,X} }$, where the random variable~$X$ is distributed
according to~$p$, and~$p$ is said to be~$\delta$-biased if its bias
with respect to every non-zero string~$u$ is bounded by~$\delta$.
\end{definition}
The bias with respect to a string~$u$ is the bias of the XOR
(exclusive OR) of the bits selected by the string~$u$, i.e., the
difference of the probabilities that this XOR is~$0$ or~$1$. The set
of all strings has bias zero, and small-bias spaces are more efficient
substitutes for this set.

Recall from equation~(\ref{eqn-paulimap}) in Section~\ref{sec-prelim}
that a subset of strings~$S \subset \set{0,1}^{2n}$ defines a CPTP map
on~$n$ qubits as follows:
\begin{eqnarray}
\label{def-map}
R_S(\rho) & = & \frac{1}{\size{S}} \sum_{(a,b)\in S} 
              \paulix^a \pauliz^b \rho \pauliz^b \paulix^a.
\end{eqnarray}
If we choose~$S = \set{0,1}^{2n}$, we get a completely randomizing
map. Ambainis and Smith showed that if we choose~$S$ to be
a~$\delta$-biased set, then the operator~$R_S$ scales every
non-identity Pauli operator by a factor at most~$\delta$. We then get
an~$\epsilon$-randomizing map by setting~$\delta$ to be suitably
small, namely, $\epsilon \cdot 2^{-n/2}$.
\begin{proposition}[Ambainis and Smith~\cite{AmbainisS04}]
\label{thm-as}
Let~$S \subset \set{0,1}^{2n}$ be a set with bias at
most~$\epsilon/2^{n/2}$. Then the map~$R_S$ as defined in
equation~(\ref{def-map}) is an~$\epsilon$-randomizing map with respect
to the trace norm for~$n$-qubit states.
\end{proposition}
For completeness, we give a proof of this proposition in
Appendix~\ref{sec-proofs}.

We now use an optimal construction of~$\delta$-biased sets to get our
randomizing map.
\begin{proposition}[Alon, Goldreich, H{\aa}stad, Peralta~\cite{AlonGHP92}]
\label{thm-smallbias}
Let~$r,s$ be positive integers. There is a subset~$S \subset
\set{0,1}^{rs}$, of size~$2^{2r}$, with bias at
most~$\frac{s}{2^r}$. Given a monic irreducible polynomial of
degree~$r$ over~$\GF(2)$, and an index~$1 \leq i \leq rs$, the~$i$-th
string in~$S$ may be constructed with~$\Order(rs)$ multiplications
in~$\GF(2^r)$, and a further~$r^2 s$ bit operations.
\end{proposition}
We describe this construction in Appendix~\ref{sec-smallbias}.

For our purposes, we need~$r,s$ such that the length of the strings
is~$2n$, and the bias of the set~$S$ is at most~$\epsilon \cdot
2^{-n/2}$. In other words,
\begin{eqnarray*}
rs & = &  2n, \\
\frac{s}{2^{r}} & \leq & \epsilon \cdot 2^{-n/2}.
\end{eqnarray*}
Solving for the smallest such~$r$, we get that the length of key~$2r$ is
at most
\[
2r \quad \leq \quad \left\lceil n + 2 \log\frac{1}{\epsilon} + 4
\right\rceil.
\]
So a~$\delta$-biased set of size~$m = 2^{2r} \leq 16 \cdot
2^n/\epsilon^2$ with~$\delta \leq \epsilon \cdot 2^{-n/2}$
exists. This gives us an~$\epsilon$-randomizing map~$R_S$ with~$m$
unitary operations, corresponding to a key length of~$2r$, as above.

Since a completely randomizing map exists with~$2^{2n}$ unitaries, we
may assume that~$\epsilon \geq 2^{-n/2}$ in our construction. In other
words, we may assume that~$r \leq n$.

Given a key of length~$2r$, and an irreducible polynomial of
degree~$r$ over~$\GF(2)$, the associated tensor product of single
qubit Pauli operators may be computed with~$\Order(rs) = \Order(n)$
multiplications in~$\GF(2^r)$, and a further~$\Order(r^2 s) =
\Order(n^2)$ bit operations. Multiplication in~$\GF(2^r)$ can be
implemented with~$\Order(r\log r) = \Order(n \log n)$ bit operations
(see, e.g., Theorem~8.7 and its corollary on page~288, Chapter~8, in
Ref.~\cite{AhoHU74}). The bit-complexity of these computations is
therefore~$\Order(n^2 \log n)$. Furthermore, a monic irreducible
polynomial of degree~$r$ over~$\GF(2)$ may be computed by a
deterministic algorithm that takes~$\tOrder(r^4)$ bit
operations~\cite[page~40, Theorem~3.6]{Shoup89}. Thus, this part of
the construction dominates the time complexity, which is in
effect~$\tOrder(n^4)$. These observations conclude the proof of
Theorem~\ref{thm-constructive}.

\section{The abundance of randomizing maps}
\label{sec-random}

In this section, we prove Theorem~\ref{thm-randomized}, which states
that there is a plethora of randomizing maps that use essentially the
same number of bits of key as in the explicit construction.  We use a
probabilistic argument that is similar in structure to that of Hayden
\etal.~\cite{HaydenLSW04}.  To show that~$m$ unitaries suffice, we
first show that a sequence of~$m$ random unitary operations
approximately randomize any fixed state with high probability. To
extend the approximate randomizing property to all states, we show
that it suffices to randomize a set of {\em finitely many\/} pure
states that in a certain precise sense approximately cover the unit
sphere in~$\complex^d$.  Finally, a ``union bound'' shows that with
probability exponentially close to~$1$ {\em every\/} state is
approximately randomized.

In our argument, each unitary operator is independently distributed
according to the Haar measure, or any other distribution over unitary
operations corresponding to a completely randomizing map.  In
particular, the operators could be chosen uniformly at random from an
orthogonal basis for~$\linear(\complex^{2^n})$, such as the Pauli
basis~$\pauli_n$.

\begin{proof}%
(of Theorem~\ref{thm-randomized}) 
Consider a sequence of~$m$ unitaries~$\set{U_i}$ independently chosen
from a measure~$\mu_i$ on~$\unitary(\complex^d)$. We require that the
measure~$\mu_i$ give us a completely randomizing map. For any density
matrix~$\rho \in \linear(\complex^d)$, and~$U$ distributed according
to~$\mu_i$,
\begin{eqnarray}
\label{eqn-mu}
\expct_U \; U\rho U^\adjoint ~~=~~ 
\integral U \rho U^\adjoint \, d\mu_i & = & \frac{\identity}{d}.
\end{eqnarray}

The sequence~$\set{U_i}$ define the map
\begin{eqnarray*}
R(\rho) & = & \frac{1}{m} \sum_{i = 1}^m U_i \rho U_i^\adjoint.
\end{eqnarray*}
Fix a pure state~$\rho \in L(\complex^d)$. We first bound the expected
distance of~$R(\rho)$ from the completely mixed
state~$\identity/d$. 
Define the random variable~$Y_\rho$ as follows
\begin{eqnarray*}
Y_\rho & = & \trnorm{ R(\rho) - \frac{\identity}{d} }.
\end{eqnarray*}
While we may carry out a similar analysis for a mixed state~$\rho$, it
is sufficient (and also simpler) to restrict ourselves to pure states;
cf.\ Remark~\ref{rem-pure}.

\begin{proposition}
\label{thm-ey}
$\expct\, Y_\rho  ~~\leq~~ \sqrt{d/m}$.
\end{proposition}
\begin{proof}
From Corollary~\ref{thm-norms}, we have
\begin{eqnarray}
\label{eqn-y}
Y_\rho^2 & \le & d \; \norm{ R(\rho)}_F^2 - 1.
\end{eqnarray}
By the definition of Frobenius norm,
\begin{eqnarray}
\norm{R(\rho)}_F^2
  & = & \trace\;  R(\rho)^2 \nonumber \\
  & = & \frac{1}{m^2} \sum_i \trace \left( U_i \rho U_i^\adjoint \right)^2
        + \frac{1}{m^2} \sum_{i \not= j} \trace\left( U_i \rho U_i^\adjoint 
          U_j \rho U_j^\adjoint \right) \nonumber \\
\label{eqn-r}
  & = & \frac{1}{m} + \frac{1}{m^2} \sum_{i \not= j} 
          \trace\left( U_i \rho U_i^\adjoint U_j \rho U_j^\adjoint \right).
\end{eqnarray}
Here, we have used the linearity of the trace function, and the fact
that~$\trace(\sigma^2) = 1$ for any pure state density
matrix~$\sigma$.

Recall that the unitary operators~$U_i$ are chosen randomly according
to a measure~$\mu_i$ satisfying equation~(\ref{eqn-mu}).  Taking
expectation over the random choice of unitaries, we get
\begin{eqnarray}
\expct_{\set{U_i}} [ \trace \,  R(\rho)^2 ]
  & = & \frac{1}{m} +  \frac{1}{m^2} \sum_{i \not= j} \expct_{\set{U_i}} 
          \trace\left( U_i \rho U_i^\adjoint U_j \rho U_j^\adjoint \right)
          \nonumber \\ 
\label{eqn-er}
  & = & \frac{1}{m} + 
          \trace\left[  (\expct_{U_i} U_i \rho U_i^\adjoint) 
          (\expct_{U_j} U_j \rho U_j^\adjoint) \right] \\
  & =   & \frac{1}{m} + \trace\; \frac{\identity}{d^2} \nonumber \\
\label{eqn-expct}
  & =   & \frac{1}{m} + \frac{1}{d}.
\end{eqnarray}
In equation~(\ref{eqn-er}), we used the fact that~$U_i$ and~$U_j$ are
chosen independently according
to measures~$\mu_i,\mu_j$.

Putting equations~(\ref{eqn-y}) and~(\ref{eqn-expct}) together gives
us 
\begin{eqnarray*}
\expct \, Y_\rho & \leq & \sqrt{\expct \,  Y_\rho^2} \\
    & \leq & \sqrt{d \, \norm{ R(\rho) }_F^2 - 1} \\
    & = & \sqrt{d/m}.
\end{eqnarray*}
the claimed bound on~$\expct\; Y_\rho$.
\end{proof}

Thus, the random sequence of unitary operators~$\set{U_i}$ randomizes
any {\em fixed\/} state~$\rho$ very well in expectation, provided~$m$
is chosen suitably larger than~$d$.

We now note that the function~$f_\rho(U_1,U_2,\ldots,U_m)$ defining the
random variable~$Y_\rho$ has bounded difference. In
other words, if we replace any one of the unitaries~$U_i$ by another
unitary~$\tilde{U}_i$, the function value changes by a small amount.
Denote the randomizing map given by the modified
sequence
$$(U_1,U_2,\ldots,U_{i-1},\tilde{U}_i,U_{i+1}, \ldots, U_m)$$
by~$\tilde{R}$. Then, we have
\begin{align}
\lefteqn{\size{f_\rho(U_1,U_2,\ldots,\tilde{U}_i,\ldots,U_m)
   - f_\rho(U_1,U_2,\ldots,U_m)} } \\
  & =~~ \size{ \trnorm{R(\rho) - \frac{\identity}{d} }
        - \trnorm{\tilde{R}(\rho) - \frac{\identity}{d} } } 
        & \notag \\
  & \le~~ \trnorm{ R(\rho) - \tilde{R}(\rho) }, & \textrm{By the
        triangle inequality} \notag \\
  & =~~ \frac{1}{m} \trnorm{ U_i \rho U_i^\adjoint - \tilde{U}_i \rho
        \tilde{U}_i^\adjoint } & \notag \\
  & \le~~ \frac{2}{m}. \label{eqn-ck} & 
\end{align}

The McDiarmid bound from probability theory states that any random
variable with such a bounded difference property is concentrated
around its mean.
\begin{theorem}[McDiarmid's Inequality~\cite{McDiarmid89}]
\label{thm-mcdiarmid}
Let~$X_1,X_2,\ldots,X_m$ be~$m$ independent random variables,
with~$X_k$ taking values in a set~$A_k$ for each~$k$. Suppose that the
measurable function~$f : \prod_{i = 1}^{m} A_i \rightarrow \reals$ satisfies
\begin{eqnarray*}
\size{f(x) - f(x')} & \le & c_k
\end{eqnarray*}
whenever the vectors~$x$ and~$x'$ differ only in the~$k$-th
coordinate. Let~$Y = f(X_1,X_2,\ldots,X_m)$ be the corresponding random
variable. Then for any~$t \ge 0$,
\begin{eqnarray*}
\Pr\left[ Y - \expct(Y) \ge t \right] 
    & \le &  \exp\left( \frac{- 2 t^2}{\sum_{i=1}^m c_k^2} \right) 
\end{eqnarray*}
\end{theorem}

Theorem~\ref{thm-mcdiarmid}, along with equation~(\ref{eqn-ck})
immediately implies that for any fixed pure state~$\rho \in
\complex^d$,
\begin{eqnarray*}
\Pr[Y_\rho - \expct\, Y_\rho \ge \delta] 
& \le & \exp\left( \frac{ - \delta^2 m}{2} \right).
\end{eqnarray*}
This implies, using our bound from Proposition~(\ref{thm-ey}) on the
expected value of~$Y_\rho$,
\begin{eqnarray}
\label{eqn-tailbound}
\Pr[Y_\rho \ge \delta + \sqrt{d/m}] 
& \le & \exp\left( \frac{- \delta^2 m}{2} \right).
\end{eqnarray}

The probability that~$R(\rho)$ deviates from the completely mixed
state decays exponentially in its distance, and the number of unitary
operators~$m$.  We would like to extend this property to {\em all\/}
pure states. For this, it suffices to randomize a suitably large, but
finite, set of pure states (a ``net'') given by the following
proposition (see, e.g.\ Ref.~\cite{HaydenLSW04} for a proof).
\begin{proposition}
\label{thm-cover}
For every~$0 < \eta < 1$, there is a set~$\emm$ of pure states
in~$\complex^d$ with~$\size{\emm} \le (5/\eta)^{2d}$, such that for
every pure state~$\ket{\phi} \in \complex^d$, there is a
state~$\ket{\tilde{\phi}} \in \emm$ with~$\trnorm{\density{\phi} -
\density{\tilde{\phi}}} \le \eta$.
\end{proposition}

From Proposition~\ref{thm-cover}, we know that every pure state~$\rho
\in \complex^d$ is~$\eta$-close in trace norm to a pure 
state~$\tilde{\rho}$ from a finite set~$\emm$
of size~$\size{\emm} \le \left(\frac{5}{\eta}\right)^{2d}$. By the
triangle inequality, and the unitary equivalence of the trace norm,
it is straightforward to show that~$\size{Y_\rho -
Y_{\tilde{\rho}}} \le \eta$. Therefore, if~$Y_\rho \ge \epsilon$,
then~$Y_{\tilde{\rho}} \ge \epsilon - \eta$ for some~$\tilde{\rho} \in
\emm$.

We can now bound the probability that the map~$R$ fails to randomize
some pure state.
\begin{align*}
\lefteqn{ \Pr\left[ \exists \rho : Y_\rho > \epsilon \right] } \\
  & \le~~ \Pr\left[ \exists \tilde{\rho} \in \emm :
          Y_{\tilde{\rho}} > \epsilon - \eta \right]
          & \textrm{From the discussion above} \\
  & \le~~ \size{\emm} \cdot \Pr \left[ Y_{\tilde{\rho}} 
                                > \epsilon - \eta \right]
          & \textrm{By the union bound, for the worst case state } 
            \tilde{\rho} \\
  & \le~~ \left(\frac{5}{\eta}\right)^{2d} 
          \exp\left( \frac{-m}{2} (\epsilon - \eta - \sqrt{d/m}\,)^2 \right)
          & \textrm{By equation~(\ref{eqn-tailbound})} \\
  & \le~~ \e^{-d/2}, & 
\end{align*}
if~$\eta$ is chosen to be at most~$\epsilon/3$, and~$m$ at least
\[
\frac{37d}{\epsilon^2} \ln \left( \frac{15}{\epsilon} \right).
\]
Thus, there is an overwhelming majority of~$m =
\Order(\tfrac{d}{\epsilon^2} \log \tfrac{1}{\epsilon})$ unitaries such
that the corresponding map is randomizing to within~$\epsilon$, with
respect to the trace norm.
\end{proof}

\section{Towards proving optimality}
\label{sec-lowerbd}

The best known lower bound for an $\epsilon$-randomizing map~$R$ with
respect to the trace norm, defined by a distribution over unitary maps
on~$n$ qubits, is~$n + \log(1 - \tfrac{\epsilon}{2})$. This follows
directly from a rank argument: consider the image of a pure state. It
has rank at most~$m$, the number of unitary matrices defining~$R$. The
distance of any rank~$m$ density matrix from the completely mixed
state~$\identity/2^n$ is at least~$2(1 - m\cdot 2^{-n})$. Since~$R$ is
$\epsilon$-randomizing, this distance is at most~$\epsilon$, and the
bound on the number of bits of key, which is~$\log m$, follows.

The above lower bound does not reflect the amount of key required to
achieve better security, as~$\epsilon \rightarrow 0$. (At~$\epsilon =
0$, the optimal number of key bits is~$2n$.)  To get stronger bounds,
we focus on the simplest and perhaps most natural maps, those defined
by distributions of Pauli operators, as in
equation~(\ref{eqn-paulimap}).

Recall that the~$n$-qubit Pauli operators are in one-to-one
correspondence with the set~$\set{0,1}^{2n}$, and we may therefore
study distributions on this set instead. We derive conditions on these
distributions (stated in Theorem~\ref{thm-constraints}) which we 
believe will help prove the optimality of our constructions. As a
corollary, we prove that any distribution corresponding to
an~$\epsilon$ randomizing map (with respect to the trace norm) is
necessarily~$\epsilon$-biased (cf.\ Definition~\ref{def-bias}). This
implies a new lower bound on the number of bits of key. The bound
makes the strong dependence of key length on the parameter~$\epsilon$
explicit, while sacrificing the strong dependence on message
length~$n$.

In Theorem~\ref{thm-constraints}, we stated constraints on
distributions over Pauli matrices that are randomizing. We prove these
constraints here by considering the action of randomizing maps on
stabilizer states.

\begin{proof} (of Theorem~\ref{thm-constraints})
Let~$\ket{\psi}$ be an $n$-qubit (pure) stabilizer state, stabilized
by a group whose~$n$ generators are given by the set~$T$. We claim
that for any Pauli operator~$P$, the state~$P\ket{\psi}$ is either
parallel to~$\ket{\psi}$ or orthogonal to it.

If~$P$ commutes with every Pauli operator in~$T$, then~$P\ket{\psi}$
is also stabilized by~$T$: For~$g \in T$, we have~$gP\ket{\psi} =
Pg\ket{\psi} = P\ket{\psi}$. Since the linear subspace stabilized
by~$T$ is one-dimensional,~$P\ket{\psi}$ belongs the linear span
of~$\ket{\psi}$. If~$P$ anticommutes with some~$g \in T$,
then~$\bra{\psi}P\ket{\psi} = \bra{\psi} Pg\ket{\psi} = - \bra{\psi}
gP\ket{\psi} = - \bra{\psi}P\ket{\psi} = 0$.

It follows that for any two Pauli operators~$P,Q$, the
states~$P\ket{\psi}$ and~$Q\ket{\psi}$ are either parallel or
orthogonal---we use the matrix~$PQ = P^\adjoint Q$ in the above
argument. In fact, we can say something stronger.  Let~$M$ be the~$n
\times 2n$ matrix representation of the generator set~$T$. The
states~$P\ket{\psi}$ and~$Q\ket{\psi}$ are parallel iff $M \cdot w = M
\cdot w'$, where~$w$ and~$w'$ are the~$2n$-bit representations of the
Pauli operators~$P$ and~$Q$, respectively, and~$M \cdot z$ is the
vector of symplectic inner products of the rows of~$M$ with~$z$. This
is because
\[
M \cdot w = M \cdot w' \quad \text{iff} \quad M \cdot (w + w') = 0,
\]
which is equivalent to saying that~$PQ$ commutes with the stabilizer.

Let~$\ket{\psi_{x}}$ be a canonical pure state in the linear span
of~$P\ket{\psi}$, where~$P$ is any Pauli matrix such that~$M \cdot w =
x$, and~$w \in \set{0,1}^{2n}$ represents~$P$. Since the~$n$
generators in~$T$ are independent, the matrix~$M$ has
rank~$n$. Therefore, the image of set~$\set{0,1}^{2n}$ under~$M$ is
all of~$\set{0,1}^n$, the states~$\ket{\psi_x}$ are well-defined
as~$x$ ranges in~$\set{0,1}^n$, and they form an orthonormal basis
for~$\complex^{2^n}$.

Now consider a randomizing map~$R_p$ specified by a distribution~$p$
over~$\set{0,1}^{2n}$, and its action on the stabilizer
state~$\ket{\psi}$. We use~$\psi, \psi_x$, etc.\ as shorthand for the
density matrices~$\density{\psi}, \density{\psi_x}$, etc.
\begin{eqnarray*}
R_p(\psi) & = & \sum_{(a,b) \in \set{0,1}^{2n}}
                p(a,b) \; \paulix^a \pauliz^b \psi  \pauliz^b \paulix^a \\
          & = & \sum_{(a,b)} p(a,b) \; \psi_{M \cdot (a,b)}.
\end{eqnarray*}
This mixed state is diagonal in the basis~$\set{\ket{\psi_x}}$, and
therefore its trace distance from the completely mixed state is
\begin{eqnarray*}
\trnorm{ R_p(\psi) - \frac{\identity}{2^n} }
    & = & \sum_{x \in \set{0,1}^n} 
          \size{ \Pr[M \cdot V = x] - \frac{1}{2^n} },
\end{eqnarray*}
where~$V$ is the random variable corresponding to the
distribution~$p$. Since~$R_p$ is~$\epsilon$-randomizing, the above
expression is bounded by~$\epsilon$. This is precisely the~$\ell_1$
distance of the random variable~$M \cdot V$ from uniform on~$n$-bits.
\end{proof}

These conditions imposed by on distributions over Pauli matrices are
similar to conditions such as ``almost $k$-wise independence'' (see,
e.g., Ref.~\cite{NaorN93}), but are not equivalent to any of the
standard notions of pseudo-randomness. As claimed in
Corollary~\ref{thm-converse}, it is however a stronger notion than
that of having bias at most~$\epsilon$. We finish with a proof of this
corollary, which also gives us a stronger lower bound for the key size
for exponentially small~$\epsilon$.

\begin{proof} (of Corollary~\ref{thm-converse})
Consider any non-zero string~$w \in \set{0,1}^{2n}$.  Let~$w = (u,v)$,
where~$u,v \in \set{0,1}^n$.  We would like to show that the random
bit~$\dotprod{w,V}$ has bias at most~$\epsilon$, where~$V$ is the
random variable corresponding to the distribution~$p$.

We first prove this property for~$w$ such that for each~$i =
1,\ldots,n$, at least one of~$u_i,v_i$ is~$1$.  Consider~$n$
stabilizer generators, the~$i$-th one~$g_i$ defined as~$g_i =
\Tensor_{j = 1}^n P_j$, where~$P_j = \paulii$ for all~$j \not= i$,
and~$P_i$ is equal to
\begin{eqnarray*}
\pauliz & & \text{if } u_i = 1 \not= v_i, \\
\paulix & & \text{if } v_i = 1 \not= u_i, \text{ and }\\
\pauliy & & \text{if } u_i = 1 = v_i.
\end{eqnarray*}
These~$n$ generators~$\set{g_i}$ commute and are independent, and
therefore specify a pure stabilizer state. This state is a tensor
product of~$n$ single qubit Pauli eigenvectors,
\[
\ket{0}, \quad \frac{1}{\sqrt{2}} ( \ket{0} + \ket{1} ), \quad \text{ or }
\frac{1}{\sqrt{2}} ( \ket{0} + \complexi \ket{1} ),
\]
depending upon whether the~$i$-th generator~$g_i$ has~$\pauliz,
\paulix$, or~$\pauliy$, respectively, in its~$i$-th tensor factor.

For~$i = 1,\ldots,n$, let~$e_i$ be the~$n$-bit string which is zero in
all positions except the~$i$-th. Then the~$2n$-bit string representing
the generator~$g_i$ is
\begin{eqnarray*}
\label{eqn-stab}
g_i & \leftrightarrow & (\dotprod{e_i,v}\, e_i, \dotprod{e_i,u}\,
e_i).
\end{eqnarray*}

Consider the action of the map~$R_p$ on this stabilizer state. From
Theorem~\ref{thm-constraints}, we get that the random variable~$M
\cdot V$ is~$\epsilon$ close to uniform on~$n$-bits, where~$M$ is the
matrix representing the stabilizer~$\set{g_i}$. Its rows are given by
the equation~(\ref{eqn-stab}). Note that~$M \cdot V$ is the sequence
of~$n$ bits~$u_i V_i + v_i V_{n+i} \pmod{2}$.
\suppress{
\begin{eqnarray*}
V_i & & \text{if } u_i = 1 \not= v_i, \\
V_{n+i} & & \text{if } v_i = 1 \not= u_i, \\
V_i \xor V_{n+i} & & \text{if } u_i = 1 = v_i.
\end{eqnarray*}}
Any distribution that is~$\epsilon$ close to uniform in~$\ell_1$-norm
is also~$\epsilon$-biased. Therefore, the XOR of the bits in~$M \cdot
V$ has bias at most~$\epsilon$. The XOR is precisely the scalar
product~$\dotprod{(u,v),V} = \dotprod{w,V}$, so we have proven the
first part of the claim for strings~$w$ of the type described above.

For an arbitrary non-zero string~$w = (u,v)$, we consider a string~$w'
= (u',v)$ such that~$u'_i = 1$ for all~$i$ such that~$u_i = v_i = 0$,
and~$u'_i = u_i$ for the remaining~$i$. From the argument above, we
have that~$M' \cdot V$ is close to uniform, where~$M'$ is defined by
the string~$w'$. The scalar product~$\dotprod{w,V}$ is the XOR of a
{\em subset\/} of the bits in~$M' \cdot V$. Therefore its bias is also
at most~$\epsilon$.

When~$p$ is uniform over a subset~$S \subset \set{0,1}^{2n}$, we get
that the set is~$\epsilon$-biased, and the stated lower bound on its
size is given in Ref.~\cite[equation~(3), page~13]{AlonGHP92}. The same
lower bound holds for possibly non-uniform distributions~$p$ with support
on the subset~$S$~\cite{Alon-eps-bias}.
\end{proof}

\suppress{
\section{Conclusion}
\label{sec-conc}

[Look up quant-ph for other work on private quantum channels]

}

\subsection*{Acknowledgements}

We thank Debbie Leung, Harold Ollivier, and David Woodruff
for helpful discussions, Noga Alon for providing a copy
of~\cite{Alon-eps-bias},
and Alfred Menezes and Jeff Shallit for pointers to the
literature.

\appendix

\section{Proofs of some claims}
\label{sec-proofs}

In this section of the Appendix, we present proofs of some statements
made in the article.

We use the following relation between trace norm and Frobenius norm,
which is essentially an application of the Cauchy-Schwartz inequality.
\begin{proposition}
\label{thm-trf}
For any rank~$d$ matrix~$M$, $\trnorm{M} \le \sqrt{d} \cdot
\norm{M}_F$.
\end{proposition}
We use this relation in the following form.
\begin{corollary}
\label{thm-norms}
Let~$M \in \linear(\complex^d)$ be a density matrix. Then, its trace
distance from the completely mixed state~$\identity/d$ is bounded as
\[
\trnorm{M - \frac{\identity}{d}}^2 ~~\le~~ d \norm{M}_F^2 - 1.
\]
\end{corollary}
\begin{proof}
By the definition of Frobenius norm in terms of the trace inner
product,
\begin{eqnarray*}
\lefteqn{ \norm{ M - \frac{\identity}{d} }_F^2 } \\
    & =  & \norm{M}_F^2 - 2 \; \trace\, \frac{M}{d}
           + \norm{\frac{\identity}{d}}_F^2 \\
    & =  & \norm{M}_F^2 - \frac{2}{d} \; \trace \, M
           + \trace\, \frac{\identity}{d^2}  \\
    & =  & \norm{M}_F^2 - \frac{1}{d}.
\end{eqnarray*}
The corollary now follows from Proposition~\ref{thm-trf}.
\end{proof}

We can now prove Proposition~\ref{thm-as}.

\begin{proof}(of Proposition~\ref{thm-as})
First, we express a state~$\rho$ in the Pauli basis:
\[
\rho \quad = \quad \frac{1}{2^n} \sum_{(u,v) \in \set{0,1}^{2n}}
\alpha_{uv} \, \paulix^u \pauliz^v,
\]
where~$\alpha = (\alpha_{uv}) \in \complex^{2^{2n}}$
with~$\norm{\alpha}_2^2 \leq 2^n$.

Since~$\paulix$ and~$\pauliz$ anti-commute,
\begin{eqnarray*}
R_S(\paulix^u \pauliz^v)
    & = & \frac{1}{\size{S}} \sum_{(a,b) \in S}
          \paulix^a \pauliz^b \; (\paulix^u \pauliz^v) \;
          \pauliz^b \paulix^a \\
    & = & \frac{1}{\size{S}} \sum_{(a,b) \in S}
          (-1)^{\dotprod{u,b} + \dotprod{v,a}} \; \paulix^u \pauliz^v \\
    & = & \delta_{v,u} \, \paulix^u \pauliz^v,
\end{eqnarray*}
where~$\dotprod{x,y}$ is the standard scalar product of two strings
over~$\GF(2)$, and~$\delta_{v,u} \in \reals$ is given by the equation
above. Note that $\size{\delta_{v,u}} = \bias(S,(v,u))$. Thus, if~$S$
is~$\delta$-biased, then each non-identity component of any density
matrix will be scaled by a factor of~$\delta$:
\begin{eqnarray*}
R_S(\rho)
    & = & \frac{1}{2^n} \sum_{(u,v) \in \set{0,1}^{2n}} \alpha_{uv} \,
          R_S( \paulix^u \pauliz^v ) \\
    & = & \frac{1}{2^n} \sum_{(u,v)} \alpha_{uv} \, \delta_{v,u} \,
          \paulix^u \pauliz^v,
\end{eqnarray*}
where~$\size{\delta_{v,u}} = \bias(S,(v,u)) \leq \delta =
\epsilon/2^{n/2}$, for all~$(v,u) \not= 0^{2n}$. The Frobenius norm of
the randomized state is thus concentrated in the first term, the
completely mixed state.
\begin{eqnarray*}
\norm{R_S(\rho)}_F^2
    & = &    \frac{1}{2^{2n}} \sum_{(u,v)}
             \size{\alpha_{uv}}^2 \cdot \delta_{v,u}^2 \cdot 
             \norm{\paulix^u \pauliz^v}_F^2 \\
    & \leq & \frac{1}{2^{2n}} \left( 2^n + \sum_{(u,v) \not= 0^{2n}} 
             \size{\alpha_{uv}}^2 \cdot \delta^2  \cdot 2^n \right) \\
    & \leq & \frac{1}{2^n}(1 + \epsilon^2).
\end{eqnarray*}
Here, we used the bound of~$2^n$ on~$\norm{\alpha}_2^2$.

From Corollary~\ref{thm-norms},
\begin{eqnarray*}
\trnorm{R_S(\rho)}^2 & \leq & 2^n \norm{R_S(\rho)}_F^2 - 1 \\
    & \leq & \epsilon^2,
\end{eqnarray*}
and Proposition~\ref{thm-as} now follows.
\end{proof}

\section{A construction of small bias sets}
\label{sec-smallbias}

In this section, we present the construction described in
Proposition~\ref{thm-smallbias} of small-bias spaces due to Alon,
Goldreich, H{\aa}stad, and Peralta~\cite[Section~5]{AlonGHP92} (see
the remarks at the end of the Section~5 in the reference). This
construction is optimal in the regime of extremely small biases that
we are interested in.

Let~$r,s$ be positive integers.  We would like to identify a set~$S
\subset \set{0,1}^{rs}$ of size~$2^{2r}$ and with bias at most~$s
2^{-r}$. We construct~$S$ by describing a string~$s_{xy}$ for each
pair of string~$x,y \in \set{0,1}^r$. We identify both~$x$ and~$y$
with elements of the vector space~$\GF(2^r)$ over the field~$\GF(2)$
in the natural way. Let~$\set{e_i}, i \in [r] = \set{0,1,\ldots, r-1}$
be a basis for the vector space~$\GF(2^r)$.

We define the string~$s_{xy}$ bit-by-bit. For~$i \in [r], j \in [s]$,
the~$(i,j)$-th bit of~$s_{xy}$ is given by~$\dotprod{e_i x^j, y}$.
All multiplications in the expression~$e_i x^j$ are in the
field~$GF(2^r)$, and~$\dotprod{\cdot,\cdot}$ is the standard scalar
product in~$\GF(2)$. The string~$s_{xy}$ is thus given by the
following array of bits:
\[
\begin{array}{cccc}
\dotprod{e_0, y} & \dotprod{e_0 x, y} & \cdots 
    & \dotprod{e_0 x^{s-1}, y} \\
\dotprod{e_1, y} & \dotprod{e_1 x, y} & \cdots 
    & \dotprod{e_1 x^{s-1}, y} \\
 & & \vdots & \\
\dotprod{e_{r-1}, y} & \dotprod{e_{r-1} x, y} & \cdots 
    & \dotprod{e_{r-1} x^{s-1}, y}
\end{array}
\]

Note that computing all the~$rs$ bits of~$s_{xy}$ takes~$\Order(rs)$
multiplications in~$\GF(2^r)$, and a further~$\Order(r^2 s)$ bit
operations to compute scalar products.

It only remains to argue that the bias of the set~$S = \set{ s_{xy} }$
so constructed has bias at most~$s 2^{-r}$. 
\begin{proposition}
The set~$S = \set{ s_{xy} } \subset \set{0,1}^{rs}$ constructed as
above has bias at most~$\tfrac{s-1}{2^r}$.
\end{proposition}
\begin{proof}
Let~$u \in \set{0,1}^{rs}$ be any non-zero string. For any
string~$s_{xy} \in S$, we have
\begin{eqnarray*}
\dotprod{u, s_{xy}} 
    & = & \sum_{i \in [r], j \in [s]} u_{ij} \dotprod{e_i x^j, y} \\
    & = & \dotprod{ \sum_{ij} u_{ij} \, e_i x^j, \; y} \\
    & = & \dotprod{ p_u(x), y },
\end{eqnarray*}
where
\[
p_u(x) \quad = \quad \sum_{j \in [s]} 
                     \left( \sum_{i \in [r]} u_{ij} e_i \right) \;
                     x^j
\]
is a polynomial in~$x$ with coefficients in~$\GF(2^r)$ and with degree
at most~$s-1$. Since~$u$ is non-zero, and~$\set{e_i}$ are linearly
independent, the polynomial~$p_u$ is not identically~$0$.

The bias of~$S$ with respect to~$u$ is then given by
\begin{eqnarray*}
\bias(S, u) 
    & = & \size{ 1 - 2 \, \expct_{xy} \dotprod{u, s_{xy}} } \\
    & = & \size{ 1 - 2 \, \Pr_{xy} [ \dotprod{u, s_{xy}} = 1 ] } \\
    & = & \size{ 1 - 2 \, \Pr_{xy} [ \dotprod{ p_u(x), y } = 1 ]}.
\end{eqnarray*}
We may estimate the above probability as
\begin{eqnarray*}
\Pr_{xy} [ \dotprod{ p_u(x), y } = 1 ] 
    & = & \Pr_{xy} [ \dotprod{ p_u(x), y } = 1 \; | \; p_u(x) \not= 0 ] 
          \cdot  \Pr_{x} [ p_u(x) \not= 0 ] \\
    & = & \frac{1}{2} \, \Pr_{x} [ p_u(x) \not= 0 ],
\end{eqnarray*}
since the scalar product of any non-zero~$p_u(x)$ with a uniformly
random~$y$ has zero bias. Putting these together, we have
\begin{eqnarray*}
\bias(S, u)
    & = & 1 - \Pr_{x} [ p_u(x) \not= 0 ] \\
    & = & \Pr_{x} [ p_u(x) = 0 ] \\
    & \leq & \frac{s-1}{2^r},
\end{eqnarray*}
since any non-zero polynomial of degree~$s-1$ has at most~$s-1$ roots
in any field.
\end{proof}

This finishes the description of the small bias set in
Proposition~\ref{thm-smallbias}.

\end{document}